\documentstyle[12pt, aps,epsf]{revtex}  
\begin{document}  
\vskip0.5cm  
\begin{center}  
{\Large \bf   Squarks Loop Corrections to the Charged Higgs Boson  
Pair Production in Photon-Photon Collisions  
 \\}  
  \vskip1cm {\large Shou Hua Zhu,   
  Chong Sheng Li and   
  Chong Shou Gao \footnote{huald@ccastb.ccast.ac.cn, csli@pku.edu.cn and gaochsh@pku.edu.cn}}  
  \vskip0.2cm  
  Department of Physics,  
  Peking University, Beijing, 100871, P. R. China \\  
  \vskip0.2cm   
  China Center of Advanced Science and Technology(World Laboratory),\\  
  P. O. Box 8730, Beijing, 100080, P. R. China \footnote{Mailing address} \\  
  \vskip0.5cm  
  {\large\bf Abstract \\[10pt]} \parbox[t]{\textwidth}{  
We calculate the squarks one-loop corrections to the cross sections of the charged Higgs boson pair production in photon-photon collisions in the minimal  supersymmetric extension of standard model(MSSM).  
In general, in case of the large splitting in the masses of left-top squark and the right-top squark  and for $\tan\beta$ near $1.5$,      
the corrections can reduce the cross sections by more than $10\%$ for a wide range of the charged Higgs boson mass, depending on the $e^+e^-$ center-of-mass energy. But in other cases, the corrections are at most only a few percent.  
Those corrections can be comparable to the $O(\alpha m_t^2/m_W^2)$ Yukawa corrections in previous literature.  
 }  
\vskip1cm   
\end{center}  
  
PACS number(s): 12.15.Lk, 12.60.Jv, 14.80.Cp  
\newpage  
\section{Introduction}  
\setcounter{equation}{0}  
In various extension of the Higgs sector of the standard model (SM), there are physical charged Higgs bosons. For example,   
the minimal supersymmetric standard model(MSSM)\cite{MSSM} with two Higgs doublets, which give masses separately to up- and down-type fermions and assure cancellation of anomalies, predicts the existence of three neutral and two charged Higgs bosons $h_0$, $H$, $A$ and  $H^\pm$. The pursuit of the Higgs bosons predicted in the SM and the MSSM is   
one of the primary goals of the present and next generation of colliders.  
The neutral Higgs bosons productions were intensively studied in the literature \cite{MSSM}. But there are relatively less studies on the charged Higgs bosons production.  
Recently, the calculations of the charged Higgs boson pair production at hadron colliders through the gluon-gluon fusion have been carried out in Ref. \cite{Jiang}, which show that this production mechanism via quark and scalar quarks loop would dominate the usual $q\bar{q}$ annihilation Drell-Yan process if there exist sufficiently heavy quarks or  
 scalar quarks, and the production cross section can reach a few $fb$. However, the Next Linear Collider (NLC) operating at a center-of-mass energy of $500 - 2000 GeV$ with the luminosity of the order of $10^{33} cm^{-2} s^{-1}$ can provide an ideal place to search for the Higgs boson, especially, it may produce  a   charged Higgs boson pair with   
much larger observable production rate than at the hadron colliders, because the production process can occur at the tree level and isn't  suppressed by the Yukawa couplings between the light quarks and the Higgs bosons, and the events would be much cleaner than at the hadron colliders,   
meanwhile the parameters of the  charged Higgs boson would be   
easier to extracted.  
In Ref. \cite{EEHPHN}, the process $e^+e^- \rightarrow H^+H^-$ and  
the fermion and sfermion one-loop corrections to the process at the NLC have been studied, and it pointed out that the corrections are typically 
around $-10\%$ in a wide range of the MSSM parameters. 
Although the tree level process $e^+e^- \rightarrow \gamma\gamma \rightarrow H^+H^-$ was also already studied  almost a fifteen years ago \cite{Bowser}, the one-loop radiative corrections to the process are studied just recently \cite{Ma}. But in Ref. \cite{Ma} the authors considered only the $O(\alpha m_t^2/m_W^2)$ Yukawa corrections  in the  
two-Higgs-doublets model (2HDM), in which the virtual particles are top and bottom quarks, and the complete   
one-loop radiative corrections to the process hasn't been calculated so far in both of the 2HDM and  the MSSM.   
  
In this paper, we present the calculations of the squarks loop corrections to the process $e^+e^- \rightarrow \gamma\gamma \rightarrow H^+H^-$ in the MSSM. Since, in general, the heavy superparticles decouple in loop contributions, the contributions from the lighter stop may be important, which would lead to observable effect.    
  The  complete one-loop supersymmetric corrections to the process should include all genuine supersymmetric virtual particles, which will be discussed in details   
elsewhere \cite{Zhu}.  
The structure of this paper is as follows. In Sec. II we give the analytical results in terms   
of the well-known standard notation of the one-loop integrals  
\cite{denner}. In Sec. III we present some   
numerical examples with discussions. The lengthy expressions of the form factors in the amplitude are summarized in Appendix.

\section{Calculations}  
  
The Feynman diagrams for the process  
\begin{eqnarray}  
\gamma (p_1,\lambda_1)+\gamma (p_2,\lambda_2)\longrightarrow H^+(k_1)+H^-(k_2),  
\end{eqnarray}  
which include the squarks loop corrections, are shown in Fig.1-4, where $\lambda_{1,2}$   
denote the helicities of photons. The tree-level   
diagrams are shown in Fig. \ref{fig1}. The self-energy, vertex and box correction diagrams  
are depicted in Fig. \ref{fig2}, \ref{fig3} and \ref{fig4}, respectively. The relevant Feynman rules can be found in Ref \cite{MSSM}. For simplicity, we define the   
Mandelstam variables as  
\begin{eqnarray}  
\hat{s}&=&(p_1+p_2)^2=(k_1+k_2)^2\nonumber \\  
\hat{t}&=&(p_1-k_1)^2=(p_2-k_2)^2 \nonumber \\  
\hat{u}&=&(p_1-k_2)^2=(p_2-k_1)^2,   
\end{eqnarray}  
and introduce the explicit   
polarization vectors of the helicities ($\lambda_1\lambda_2$) for photons as follows  
\begin{eqnarray}  
\epsilon_1^\mu (p_1, \lambda_1=\pm 1)&=&-\frac{1}{\sqrt{2}}(0,1,\mp i ,0)  \nonumber \\  
\epsilon_2^\mu (p_2, \lambda_2=\pm 1)&=&\frac{1}{\sqrt{2}}(0,1,\pm i ,0).  
\end{eqnarray}  
This choice of polarization vectors for the photons implies  
\begin{eqnarray}  
\epsilon_i.p_j =0,\ \ \ \ \ (i,j=1,2)  
\end{eqnarray}  
and by the momentum conservation,  
\begin{eqnarray}  
\epsilon_i.k_1 =-\epsilon_i.k_2,\ \ \ \ \ ( i=1,2).  
\end{eqnarray}  
  
In our calculations, we used dimensional regularization to control all the ultraviolet divergences  
in the virtual loop corrections and we adopted the on-mass-shell renormalization   
scheme \cite{denner1,Sirlin}.  
Note that dimensional reduction is usually more used in the calculations of the radiative corrections in the  MSSM as it automatically preserves supersymmetry (at least to one-loop order), while dimensional regularization does not. But  there are only virtual scalar particle in the loops in this work, and the calculations of Feynman integrals do not involve the n-dimensional Dirac algebra, so the dimensional reduction scheme is same with the conventional  dimensional regularization one here.  
Including the squarks loop corrections, the  renormalized amplitude for   
the process $\gamma\gamma\rightarrow H^+H^-$ can be written as  
\begin{eqnarray}  
M_{ren}=M_0+\delta M^{self}+\delta M^{vertex}+\delta M^{box},  
\end{eqnarray}  
where $M_0$ is the amplitude at the tree level, $\delta M^{self}$, $\delta M^{vertex}$  
and $\delta M^{box}$ represent the squarks loop corrections arising  
from the self-energy, vertex and box diagrams, respectively.  
  
The corresponding amplitude at the tree level for   
$\gamma\gamma\rightarrow H^+H^-$ is given by:  
\begin{eqnarray}  
M_0=M_0^t+M_0^u+M_0^q,  
\end{eqnarray}  
where $M_0^t$, $M_0^u$, $M_0^q$ represent the amplitude of the t-channel,  
u-channel and quadratic coupling diagrams in Fig. \ref{fig1}, respectively. Their  
explicit expressions can be given by  
\begin{eqnarray}  
M_0^t={{4\,{{{\it e}}^2}\,{\it k_1.\epsilon_1}\,{\it k_1.\epsilon_2}}\over  
       {{{{\it m_{H^\pm}}}^2} - \hat{t}}},  
\end{eqnarray}  
\begin{eqnarray}  
M_0^u={{4\,{{{\it e}}^2}\,{\it k_1.\epsilon_1}\,{\it k_1.\epsilon_2}}\over  
       {{{{\it m_{H^\pm}}}^2} - \hat{u}}},         
\end{eqnarray}  
\begin{eqnarray}  
M_0^q={{{\it e}}^2}\,{\it \epsilon_1.\epsilon_2}.  
\end{eqnarray}

The amplitude $\delta M^{self}$ arising from the self-energy corrections is given by  
\begin{eqnarray}  
 \delta M^{self}=&-&M_0^t ( {{\sum}^{H^+H^+}(\hat{t})-\delta m_{H^\pm}^2 \over \hat{t}-   
 m_{H^\pm}^2}+\delta Z_{H^\pm} )\nonumber \\   
 &-&M_0^u ( {{\sum}^{H^+H^+}(\hat{u})-\delta m_{H^\pm}^2\over \hat{u}-   
 m_{H^\pm}^2}+\delta Z_{H^\pm} ),  
\end{eqnarray}  
where ${\sum}^{H^+H^+}$, $\delta m_{H^\pm}$ and $\delta Z_{H^\pm}$ represent  the charged Higgs self-energy, the charged Higgs boson mass renormalization constant and the charged Higgs boson wave function renomalization constant, respectively,  which are given by  
\begin{eqnarray}  
{\sum}^{H^+H^+}(k^2)&=&{-N_C\over {16\,{{\pi }^2}}}\left( -i\,{\it \xi_7}\,{{{\it m_{\tilde{b}_1}}}^2} - i\,{\it \xi_8}\,  
{{{\it m_{\tilde{b}_2}}}^2} -  
        i\,{\it \xi_5}\,{{{\it m_{\tilde{t}_1}}}^2} - i\,{\it \xi_6}\,{{{\it m_{\tilde{t}_2}}}^2} \right.\nonumber\\  
        &-&  
        i\,{\it \xi_7}\,{{{\it m_{\tilde{b}_1}}}^2}\,   
         {\it B_0}(0,{{{\it m_{\tilde{b}_1}}}^2},{{{\it m_{\tilde{b}_1}}}^2}) -  
        i\,{\it \xi_8}\,{{{\it m_{\tilde{b}_2}}}^2}\,  
         {\it B_0}(0,{{{\it m_{\tilde{b}_2}}}^2},{{{\it m_{\tilde{b}_2}}}^2})\nonumber\\  
         & -&  
        i\,{\it \xi_5}\,{{{\it m_{\tilde{t}_1}}}^2}\,  
         {\it B_0}(0,{{{\it m_{\tilde{t}_1}}}^2},{{{\it m_{\tilde{t}_1}}}^2}) -  
        i\,{\it \xi_6}\,{{{\it m_{\tilde{t}_2}}}^2}\,  
         {\it B_0}(0,{{{\it m_{\tilde{t}_2}}}^2},{{{\it m_{\tilde{t}_2}}}^2}) \nonumber\\  
         &+ &  
        {{{\it \xi_1}}^2}\,{\it B_0}({k^2},{{{\it m_{\tilde{b}_1}}}^2},{{{\it m_{\tilde{t}_1}}}^2}) +  
        {{{\it \xi_4}}^2}\,{\it B_0}({k^2},{{{\it m_{\tilde{b}_1}}}^2},{{{\it m_{\tilde{t}_2}}}^2}) \nonumber \\  
&+&\left.  
        {{{\it \xi_3}}^2}\,{\it B_0}({k^2},{{{\it m_{\tilde{b}_2}}}^2},{{{\it m_{\tilde{t}_1}}}^2})   
		+  
        {{{\it \xi_2}}^2}\,{\it B_0}({k^2},{{{\it m_{\tilde{b}_2}}}^2},{{{\it m_{\tilde{t}_2}}}^2})  
         \right),   
\end{eqnarray}  
\begin{eqnarray}  
\delta m_{H^\pm}^2&=&Re{\sum}^{H^+H^+}(m_{H^\pm}^2) 
\end{eqnarray}  
\begin{eqnarray}  
\delta Z_{H^\pm}&=&-{\partial{\sum}^{H^+H^+}(k^2) \over \partial k^2} |_{k^2=  
m_{H^\pm}^2}\nonumber \\  
&=&{N_C\over {16\,{{\pi }^2}}}\left(   
        {{{\it \xi_1}}^2}\,{\it G_0}({m_{H^\pm}^2},{{{\it m_{\tilde{b}_1}}}^2},{{{\it m_{\tilde{t}_1}}}^2}) +  
        {{{\it \xi_4}}^2}\,{\it G_0}({m_{H^\pm}^2},{{{\it m_{\tilde{b}_1}}}^2},{{{\it m_{\tilde{t}_2}}}^2})   
\right. \nonumber \\   
&+&\left.  
        {{{\it \xi_3}}^2}\,{\it G_0}({m_{H^\pm}^2},{{{\it m_{\tilde{b}_2}}}^2},{{{\it m_{\tilde{t}_1}}}^2})  
		+  
        {{{\it \xi_2}}^2}\,{\it G_0}({m_{H^\pm}^2},{{{\it m_{\tilde{b}_2}}}^2},{{{\it m_{\tilde{t}_2}}}^2})  
         \right)  
\end{eqnarray}  
with  
\begin{eqnarray}  
G_0(M^2,M_1^2,M_2^2)={\partial B_0(k^2,M_1^2,M_2^2)\over \partial k^2}|_{k^2=M^2}.  
\end{eqnarray}  
Here we only explicitly   
write down the third generation scalar quarks contributions, and  
$N_C$ is the number of colors,  $\xi_i (i=1-4)$ and $\xi_i (i=5-8)$ are the coupling constants of   
the verteces of $H^+-\tilde{t}-\tilde{b}$ and $H^+-H^--\tilde{q}-\tilde{q}$,  
 respectively,   
and the explicit definitions of $\xi_i (i=1,8)$  are given  in Appendix. $B_0$ is the two-point function,  definition for which can be found in  
Ref. \cite{denner,denner1}.  
  
 The amplitude $\delta M^{vertex}$ arising from the vertex corrections is given by  
\begin{eqnarray}  
\delta M^{vertex}&=&\epsilon_1.\epsilon_2 M_{1}^{tri}+  
k_1.\epsilon_1 k_1.\epsilon_2 M_{2}^{tri}  
+  
2 (M_0^t+M_0^u)(\delta Z_e\nonumber\\  
&+&  
{1\over 2}\delta Z_{AA}+  
{\cos (2\theta_W)\over 4 \sin\theta_W\cos\theta_W} \delta Z_{ZA}+  
\delta Z_{H^\pm})  
\label{eq19}  
\end{eqnarray}  
with  
\begin{eqnarray}  
\delta Z_e={1\over 2}{\partial\sum^{AA}_{T}(k^2) \over \partial k^2} |_{k^2=0}  
-{\sin\theta_W \over \cos\theta_W }{\sum^{AZ}_{T}(0)\over M_Z^2},  
\end{eqnarray}  
\begin{eqnarray}  
\delta Z_{AA}=-{\partial\sum^{AA}_{T}(k^2) \over \partial k^2} |_{k^2=  
0},  
\end{eqnarray}  
\begin{eqnarray}  
\delta Z_{ZA}=2 {\sum^{AZ}_{T}(0)\over M_Z^2},  
\end{eqnarray}  
where the expressions of $M_{1}^{tri}$ and $M_{2}^{tri}$    
are explicitly presented in the Appendix.  
From our calculations, we find  that  
\begin{eqnarray}  
\delta Z_{ZA}=0  
\end{eqnarray}  
for squrks loop diagrams,   
and the electron charge renormalization constant $\delta Z_e$ can be canceled by the photon wavefunction renormalization constant ${1\over 2} \delta Z_{AA}$, and only $\delta Z_{H^\pm}$  remains among the renormalization constants in Eqs.   
\ref{eq19}.  
  
The amplitude $\delta M^{box}$ arising from the box corrections is given by  
\begin{eqnarray}  
\delta M^{box}&=&\epsilon_1.\epsilon_2 M_{1}^{box}+  
k_1.\epsilon_1 k_1.\epsilon_2 M_{2}^{box}  
+  
2 e^2 \epsilon_1.\epsilon_2 \delta Z_{H^\pm},  
\label{eq20}  
\end{eqnarray}  
where the explicit expressions of $M_{1}^{box}$ and  $M_{2}^{box}$  are also given in Appendix. Because of the same reason as above, we have not written  out  the $\delta Z_{ZA}$, $\delta Z_{AA}$ and $\delta Z_e$ in counterterms in Eqs.   
\ref{eq20}.  
  
The corresponding amplitude squared for the process $\gamma\gamma \rightarrow H^+H^-$  
 can be written as  
\begin{eqnarray}  
\bar{{\sum}}\left| M_{ren}\right|^2=\bar{{\sum}}\left| M_{0}\right|^2   
 +2 Re \bar{{\sum}}(\delta M^{self}+ \delta M^{vertex}+\delta M^{box}) M_0^\dagger,  
\end{eqnarray}  
where the bar over the summation recalls average over initial photons spins.  
The cross section of  process $\gamma\gamma\rightarrow H^+H^-$ is   
\begin{eqnarray}  
\hat{\sigma} =\int_{\hat{t}_{min}}^{\hat{t}_{max} }{1\over 16\pi \hat{s }^2}\bar{ \sum_{ spins} }\left|  M\right|^2 d \hat{t}  
\end{eqnarray}  
with  
\begin{eqnarray}  
\hat{t}_{min}=(m_{H^\pm}^2-{\hat{s}\over 2})-{\beta \hat{s}\over 2}\nonumber \\  
\hat{t}_{max}=(m_{H^\pm}^2-{\hat{s}\over 2})+{\beta \hat{s}\over 2},  
\end{eqnarray}  
where $\beta =\sqrt{1-4 m_{H^\pm}^2/\hat{s}} $.  
The total cross section of $e^+e^- \rightarrow \gamma\gamma \rightarrow H^+H^-$   
can be obtained by folding the $\hat{\sigma}$  with the  
photon luminosity  
\begin{eqnarray}  
\sigma (s)=\int^{x_{max}}_{2 m_{H^\pm}/\sqrt{s}} dz  
{dL_{\gamma\gamma}\over dz} \hat{\sigma} (\gamma\gamma\rightarrow H^+H^- \mbox{at  
$\hat{s}=z^2 s$}),  
\end{eqnarray}  
where $\sqrt{s}$ and $\sqrt{\hat{s}}$ is the CMS energy of $e^+e^-$ and  
$\gamma\gamma$, respectively, and $dL_{\gamma\gamma}/dz$ is the photon   
luminosity, which is defined as  
\begin{eqnarray}  
{dL_{\gamma\gamma}\over dz}=2 z \int_{z^2\over x_{max}}^{x_{max}}{dx\over x}  
f_{\gamma /e}(x) f_{\gamma /e}(z^2/x),  
\end{eqnarray}  
where $f_{\gamma /e}(x)$ is the photon structure function of the electron  
beam \cite{Halzen}. For a TeV collider with $\sigma_x/\sigma_y=25.5$, the beamstrahlung  
 photon structure function can be represented as \cite{Blank}  
\begin{eqnarray}  
f_{\gamma /e}(x)=\left\{  
\begin{array}{ll}  
\left( 2.25-\sqrt{x\over 0.166}\right)\left( {1-x\over x}\right)^{2/3}  & \mbox{ for $x<0.84$, }\\  
0  &\mbox{ for $x>0.84$},  
\end{array}  
\right.  
\end{eqnarray}  
where $x$ is the relative momentum of the radiated photon and the parent electron.  
  
If we operate NLC as a mother machine of photon-photon collider in Compton back-scattering photon fusion mode, the energy spectrum of the photons is given by \cite{Bohm}  
\begin{eqnarray}  
f_{\gamma /e}(x)=\left\{  
\begin{array}{ll}  
{1\over 1.8397}\left(1-x+{1\over 1-x}-{4x\over x_i(1-x)}+  
{4 x^2\over x_i^2(1-x)^2}\right)  & \mbox{ for $x<0.83$, $x_i=2(1+\sqrt{2})$},\\  
0  &\mbox{ for $x>0.83$}.  
\end{array}  
\right.  
\end{eqnarray}

\section{Numerical examples and conclusions}  
 
In the following we present some numerical results for the squarks loop corrections to the  
cross sections of a charged Higgs boson pair production   
in the process of   
$\gamma\gamma \rightarrow H^+H^-$ and   
$e^+e^- \rightarrow \gamma\gamma \rightarrow H^+H^-$, respectively.   
In our numerical calculations, for the SM parameters, we choose $m_w=80.33 GeV$,   
$m_z=91.187 GeV$, $m_t=176 GeV$  and $\alpha={1\over 128}$, and   
other quark masses are chosen as zero. Other parameters are determined as follows  
  
(i) The Higgs boson masses $m_{h0}$, $m_H$, $m_A$,  
$m_{H^\pm}$ and parameters $\alpha$, $\beta$ are not constrained in the general 2HDM, but  
in the MSSM, relations \cite{MSSM} among these parameters are required by SUSY, leaving only  
two parameters free, e.g., $m_{H^\pm}$ and $\tan\beta$. Also, in the MSSM the charged   
Higgs boson mass is heavier than the $W$ mass due to the relation   
$m_{H^\pm}^2=m_W^2+m_A^2$. In our numerical calculations, we will use $m_{H^\pm}$ as a variable $>100 $ GeV.  
  
(ii) In the MSSM the mass eigenstates $\tilde{q}_1$ and $\tilde{q}_2$ of the squarks are related to the current eigenstates $\tilde{q}_L$ and $\tilde{q}_R$ by \cite{MSSM}  
\begin{eqnarray}  
\left(\begin{array}{c}  
\tilde{q}_1 \\ \tilde{q}_2\end{array}\right)=  
R^{\tilde{q}}\left(\begin{array}{c}  
\tilde{q}_L \\ \tilde{q}_R\end{array}\right)\ \ \ \ \mbox{with}\ \ \ \  
R^{\tilde{q}}=\left(\begin{array}{cc}  
                   \cos\theta_{\tilde{q}} & \sin\theta_{\tilde{q}}\\  
                   -\sin\theta_{\tilde{q}} & \cos\theta_{\tilde{q}}   
                   \end{array}  
                   \right).  
\label{eq1}  
\end{eqnarray}  
For the squarks, the mixing angle $\theta_{\tilde{q}}$ and the masses $m_{\tilde  
{q}_{1,2}}$ can be calculated by diagonizing the following mass  
matrices  
\begin{eqnarray}  
M^2_{\tilde{q}}=\left(\begin{array}{cc}  
          M_{LL}^2 & m_q M_{LR}\\  
           m_q M_{RL} & M_{RR}^2  
           \end{array} \right), \nonumber \\  
M_{LL}^2=m_{\tilde{Q}}^2+m_q^2+m_{z}^2\cos 2\beta (I_q^{3L}-e_q\sin^2\theta_w),  
\nonumber \\  
M_{RR}^2= m_{\tilde{U},\tilde{D}}^2 +m_q^2+m_{z}^2\cos 2\beta e_q\sin^2\theta_w, \nonumber \\  
M_{LR}= M_{RL}=\left\{ \begin{array}{ll}  
                A_t-\mu \cot \beta & (\tilde{q}= \tilde{t})\\  
                A_b-\mu \tan \beta & (\tilde{q}= \tilde{b}),  
                \end{array}  
                \right.  
\label{eq2}  
\end{eqnarray}  
where $ m_{\tilde{Q}}^2$, $ m_{\tilde{U},\tilde{D}}^2$ are soft SUSY breaking mass terms of the left- and right-handed squark, respectively. Also, $\mu$ is the coefficient of the $H_1H_2$  
mixing term in the superpotential, $A_t$ and $A_b$ are the coefficient of the dimension-three trilinear soft SUSY-breaking term. $I_q^{3L}, e_q$ are the weak isospin and electric charge of the squark $\tilde{q}$.   
From Eqs. \ref{eq1} and \ref{eq2}, $m_{\tilde{t}_{1,2}}$ and $\theta_{\tilde{t}}$ can be derived as  
\begin{eqnarray}  
m^2_{\tilde{t}_{1,2}}&=&{1\over 2}\left[ M^2_{LL}+  
M^2_{RR}\mp \sqrt{  
(M^2_{LL}-M^2_{RR})^2+4 m^2_t M^2_{LR}}\right] \nonumber \\  
\tan\theta_{\tilde{t}}&=&{m^2_{\tilde{t}_1}-M^2_{LL} \over m_t M_{LR}}.  
\label{eq3}  
\end{eqnarray}  
In our numerical calculations, we always assume that  $m_{\tilde{U}} = m_{\tilde{D}}=m_{\tilde{Q}}$ and $M_{LR}\geq 0$.

The currently most popular supersymmetric models permit to have large splitting in the masses of the left-top squarks and right-top squarks, while the masses of all the other (left- and right- ) squarks are about the same \cite{Splitting}.  
Assuming one has a supersymmetric signal, we use the full ($>100$ parameters) parameter space freedom of the  MSSM and fit the data. This approach has been used in studying the CDF $e^+e^-\gamma\gamma+\rlap/E_T$ event \cite{16}. It was found that to also explain all the low energy data (including $\alpha_s$, $R_b$ and the branching ratio of $b \rightarrow s\gamma $, etc.), the lightest mass eigenstate ($\tilde{t}_1$) of the top squarks is about to be the order of $m_W$; $\tan\beta$ is the order of $1$; the sign ($\mu$) is negative, and $|\mu|\sim m_Z$\cite{16}. We shall refer to this class of models as MSSM models with scenarios motivated by current data.

Figure \ref{fig5}- Fig. \ref{fig7} show the relative corrections   
$\delta\hat{\sigma}/\hat{\sigma}_0$ as a function of $m_{H^\pm}$  
with $\sqrt{\hat{s}}= 500$, $1000$ and $2000$ GeV, respectively, assuming $m_{\tilde{U}} = m_{\tilde{D}}=m_{\tilde{Q}}=1$ TeV. 
The corrections in the stops mixing case can become much large 
when $\tan\beta=1.5$, which can range between $-20\%$ and $13\%$, $-20\%$ and $25\%$, $-16\%$ and $ 23\%$ for three different photon-photon CMS energy, respectively, but the corrections are all negligibly small in the no-mixing case. In order to study the variations of the corrections with respect to the $\tan\beta$, we also show the curves for $\tan\beta = 4$ and $40$, and find that the corrections are much smaller than ones for $\tan\beta=1.5$, which are at most a few percent. 
Here and below, we take the mass of lighter stop to be equal to $90$ GeV in the mixing case of stops, which is in agreement with current experiment lower limit to the squark mass \cite{ALEPH}.  
 
In our numerical calculations, we also considered the cases for the small splitting in the masses of stops, and found that the corrections are generally small except near threshold, which opens the charged Higgs boson decay into two squarks. We don't show the figures here for simplicity. 
 
Fig. \ref{fig8} gives the total cross sections as a function of charged Higgs boson masses at tree-level for $\sqrt{s}=500, 1000$ and $2000 GeV$ and two sources of photons modes. For  $m_{H^\pm}=  
150 GeV$, the total cross sections can be greater than $100 fb$ for laser  
back-scattering photons mode whether $\sqrt{s}=500, 1000 GeV$, or $ 2000 GeV$, and for $m_{H^\pm}= 300 GeV$, the total cross sections are a few tens $fb$  
and for $\sqrt{s}=1000$ and $2000$ GeV.   
But for the beamstrahlung photons mode, the total cross sections can also  
reach hundreds of $fb$ for the lighter charged Higgs boson masses for  
$\sqrt{s}=1000$ and $2000$ GeV, and they decrease rapidly with increasing the charged Higgs boson masses.

Fig. \ref{fig9} - Fig. \ref{fig11} present the relative corrections to the total cross sections as a function of $m_{H^\pm}$ with $\sqrt{s}=500,1000$ and $ 2000 GeV$, respectively, assuming $m_{\tilde{U}} = m_{\tilde{D}} = m_{\tilde{Q}} = 1$ TeV, for two sources of photon modes. 
In general, those corrections always decrease the tree-level total cross sections in the mixing case of stops, which can exceed $-10\%$ for a wide range of charged Higgs boson mass for $\tan\beta = 1.5$. However in other cases, the corrections are at most only a few percent reduction.

In conclusion, we have calculated the squarks one-loop corrections to the cross sections of the charged Higgs boson pair production in photon-photon collisions  in the MSSM.   
In general, in case of the large splitting in the masses of left-top squark and the right-top squark  and for $\tan\beta$ near $1.5$,      
the corrections can reduce the cross sections by more than $10\%$ for a wide range of the charged Higgs boson mass, depending on the CMS energy $\sqrt{s}$. But in other cases, the corrections are at most only a few percent.  
Those corrections can be comparable to the $O(\alpha m_t^2/m_W^2)$ Yukawa corrections \cite{Ma}.

\section*{Acknowledgments}  
This work was supported in part by the National Natural Science Foundation  
of China, a grant from the State Commission of Science and  
Technology of China, and the Doctoral Program 
Foundation of Institution of Higher Education, the State Education 
Commission of China. 
  
\section*{Appendix}  
In this appendix, we present the explicit analytic expressions of the form factors  
only including the third  
generation squarks contributions.  
  
$M_{1}^{tri}$ is given by  
\begin{eqnarray}  
M_{1}^{tri}=\sum_{i=1}^{12} f_{1}^{(tri, i)},  
\end{eqnarray}  
where the form factor $f_1^{(tri, i)}$ represents the contributions arising from Feynman   
diagrams with the indices of $i$ in Fig. 3.  
Here   
\begin{eqnarray}  
 f_{1}^{(tri, j)}=0  \mbox{ for $j=1, 2, 3, 4, 5, 6, 7, 8, 11, 12$},  
\end{eqnarray}  
\begin{eqnarray}  
f_{1}^{(tri, 9)}&=&{i e^2 g m_w (2 \cos(\alpha  - \beta ) - \cos(2 \beta ) \cos(\alpha  + \beta )) \over  
4 \pi^2 (\hat{s} - m_{H}^2 )}\left(\right.\nonumber\\  
&& \left[\xi_{16} e_b^2 C_{00}( 0, 0, \hat{s}, m_{\tilde{b}_2}^2, m_{\tilde{b}_2}^2, m_{\tilde{b}_2}^2)\right]  
 \nonumber \\  
&+&  
\left[ \xi_{16}\rightarrow \xi_{15}, m_{\tilde{b}_2} \rightarrow m_{\tilde{b}_1} \right]+  
\left[ e_b\rightarrow e_t, \xi_{16}\rightarrow \xi_{10}, m_{\tilde{b}_2} \rightarrow m_{\tilde{t}_2} \right]\nonumber \\  
&+&\left.  
\left[ e_b\rightarrow e_t, \xi_{16}\rightarrow \xi_{9}, m_{\tilde{b}_2} \rightarrow m_{\tilde{t}_1} \right]\right)\nonumber \\  
&+&  
{i g m_w (2 \sin(\alpha  - \beta ) - \cos(2 \beta ) \sin(\alpha  + \beta )) \over  
4 \pi^2 (m_{h0}^2 - \hat{s})}\left( \right.\nonumber\\  
&&\left[\xi_{19} e_b^2 C_{00}( 0, 0, \hat{s}, m_{\tilde{b}_2}^2, m_{\tilde{b}_2}^2, m_{\tilde{b}_2}^2)\right]  
 \nonumber \\  
&+&  
\left[ \xi_{19}\rightarrow \xi_{18}, m_{\tilde{b}_2} \rightarrow m_{\tilde{b}_1} \right]+  
\left[ e_b\rightarrow e_t, \xi_{19}\rightarrow \xi_{13}, m_{\tilde{b}_2} \rightarrow m_{\tilde{t}_2} \right]\nonumber \\  
&+&\left.  
\left[ e_b\rightarrow e_t, \xi_{19}\rightarrow \xi_{12}, m_{\tilde{b}_2} \rightarrow m_{\tilde{t}_1} \right]\right),  
\end{eqnarray}  
\begin{eqnarray}  
f_{1}^{tri, 10}&=&{i e^2 g  m_w (2 \cos(\alpha  - \beta )   
- \cos(2 \beta ) \cos(\alpha  + \beta ))\over 16 \pi^2 (m_{H}^2 - \hat{s})}  
\left(\xi_{16} e_b^2  B_0(\hat{s}, m_{\tilde{b}_2}^2, m_{\tilde{b}_2}^2)\right.  
\nonumber\\  
&+&  
\xi_{15} e_b^2 B_0(\hat{s}, m_{\tilde{b}_1}^2, m_{\tilde{b}_1}^2)   
+\left. \xi_{10} e_t^2  B_0(\hat{s}, m_{\tilde{t}_2}^2, m_{\tilde{t}_2}^2)+  
\xi_{9} e_t^2  B_0(\hat{s}, m_{\tilde{t}_1}^2, m_{\tilde{t}_1}^2)\right)\nonumber \\  
&+&  
{i g  m_w (2 \sin(\alpha  - \beta )   
- \cos(2 \beta ) \sin(\alpha  + \beta ))\over 16 \pi^2 (\hat{s}-m_{h0}^2 )}  
\left(\xi_{19} e_b^2  B_0(\hat{s}, m_{\tilde{b}_2}^2, m_{\tilde{b}_2}^2)  
\right.\nonumber \\  
&+&  
\xi_{18} e_b^2 B_0(\hat{s}, m_{\tilde{b}_1}^2, m_{\tilde{b}_1}^2)  
+\left.  \xi_{13} e_t^2  B_0(\hat{s}, m_{\tilde{t}_2}^2, m_{\tilde{t}_2}^2)+  
\xi_{12} e_t^2  B_0(\hat{s}, m_{\tilde{t}_1}^2, m_{\tilde{t}_1}^2)\right),  
\end{eqnarray}  
where $e_t$, $e_b$ are the electric charges of the top and bottom quarks, respectively. Here and  
below, $B_m$, $C_m$, $C_{mn}$, $D_m$, $D_{mn}$ are the two-, three-, four-point Feynman 
integrals, definitions of which can be found in Ref. \cite{denner,denner1}. $\xi_j, (j=9, 10)$ is the coupling constants  
of the vertex $H-\tilde{t}-\tilde{t}$, which are given by  
\begin{eqnarray}  
\left( \begin{array}{cc}  
\xi_{9} & \xi_{11} \\  
\xi_{11} & \xi_{10}  
\end{array}\right)=(-i g) R^{\tilde{t}}(A_{H\tilde{t}\tilde{t}}) (R^{\tilde{t}})^T  
\end{eqnarray}  
with  
\begin{eqnarray}  
A_{H\tilde{t}\tilde{t}}&=&  
\left(\begin{array}{cc}  
 {m_z ((1/2)-(2/3) \sin^2\theta_w) \cos ( \alpha +\beta  )\over \cos\theta_w} +  
 {m_t^2 \sin ( \alpha )\over m_w \sin ( \beta  )}  
&{m_t\over 2 m_w \sin\beta} (A_t \sin\alpha-\mu \cos\alpha) \\  
 {m_t\over 2 m_w \sin\beta} (A_t \sin\alpha-\mu \cos\alpha)&  
 {m_z (2/3) \sin^2\theta_w \cos (\alpha+\beta)\over \cos\theta_w}+  
{m_t^2 \sin\alpha \over m_w \sin\beta}  
 \end{array} \right).    
\end{eqnarray}  
$\xi_j, (j=12, 13)$ is the coupling constants  
of the vertex $h_0-\tilde{t}-\tilde{t}$, which are given by  
\begin{eqnarray}  
\left( \begin{array}{cc}  
\xi_{12} & \xi_{14} \\  
\xi_{14} & \xi_{13}  
\end{array}\right)=(i g) R^{\tilde{t}}(A_{h_0\tilde{t}\tilde{t}}) (R^{\tilde{t}})^T   
\end{eqnarray}  
with  
\begin{eqnarray}  
A_{h_0\tilde{t}\tilde{t}}  
&=& \left(\begin{array}{cc}  
 {m_z ((1/2)-(2/3) \sin^2\theta_w) \sin ( \alpha +\beta  )\over \cos\theta_w} -  
 {m_t^2 \cos ( \alpha )\over m_w \sin ( \beta  )}  
&{m_t\over 2 m_w \sin\beta} (-A_t \cos\alpha-\mu \sin\alpha) \\  
 {m_t\over 2 m_w \sin\beta} (-A_t \cos\alpha-\mu \sin\alpha)&  
 {m_z (2/3) \sin^2\theta_w \sin (\alpha+\beta) \over \cos\theta_w}-  
{m_t^2 \cos \alpha \over m_w \sin\beta}  
 \end{array} \right).    
\end{eqnarray}  
$\xi_j, (j=15, 16)$ is the coupling constants  
of the vertex $H-\tilde{b}-\tilde{b}$, which are given by  
\begin{eqnarray}  
&&\left( \begin{array}{cc}  
\xi_{15} & \xi_{17} \\  
\xi_{17} & \xi_{16}  
\end{array}\right)=(-i g)R^{\tilde{b}}(A_{H\tilde{b}\tilde{b}}) (R^{\tilde{b}})^T  
\end{eqnarray}  
with  
\begin{eqnarray}   
A_{H\tilde{b}\tilde{b}}&=&  
  \left(\begin{array}{cc}  
 {-m_z ((1/2)-(1/3) \sin^2\theta_w) \cos ( \alpha +\beta  )\over \cos\theta_w} +  
 {m_b^2 \cos ( \alpha )\over m_w \cos ( \beta  )}  
&{m_b\over 2 m_w \cos\beta} (A_b \cos\alpha-\mu \sin\alpha) \\  
 {m_b\over 2 m_w \cos\beta} (A_b \cos\alpha-\mu \sin\alpha)&  
 {-m_z\over 3 \cos\theta_w }\sin^2\theta_w \cos (\alpha+\beta)+  
{m_b^2 \cos\alpha \over m_w \cos\beta}  
 \end{array} \right).   
\end{eqnarray}  
$\xi_j, (j=18, 19)$ is the coupling constants  
of the vertex $h_0-\tilde{b}-\tilde{b}$, which are given by  
\begin{eqnarray}  
&&\left( \begin{array}{cc}  
\xi_{18} & \xi_{20} \\  
\xi_{20} & \xi_{19}  
\end{array}\right)= (i g) R^{\tilde{b}} (A_{h_0\tilde{b}\tilde{b}} )  (R^{\tilde{b}})^T   
\end{eqnarray}  
with  
\begin{eqnarray}  
A_{h_0\tilde{b}\tilde{b}}&=&  
 \left(\begin{array}{cc}  
 {-m_z ((1/2)-(1/3) \sin^2\theta_w) \sin ( \alpha +\beta  )\over \cos\theta_w} +  
 {m_b^2 \sin ( \alpha )\over m_w \cos ( \beta  )}  
&{m_b\over 2 m_w \cos\beta} (A_b \cos\alpha+\mu \sin\alpha) \\  
 {m_b\over 2 m_w \cos\beta} (A_b \cos\alpha+\mu \sin\alpha) &  
 {-m_w\over 3 \cos\theta_w}\sin^2\theta_w \sin (\alpha+\beta)+  
{m_b^2 \sin \alpha \over m_w \cos\beta}  
 \end{array} \right).   
\end{eqnarray}  
  
$M_{2}^{tri}$ is given by  
\begin{eqnarray}  
M_{2}^{tri}&=&\sum_{i=1}^{12} f_{2}^{(tri, i)},  
\end{eqnarray}  
where the form factor $f_2^{(tri, i)}$ represents the contributions arising from Feynman   
diagrams with the indices of $i$ in Fig. 3.  
Here   
\begin{eqnarray}  
f_{2}^{(tri, j)}=0  \mbox{for $j=1, 2, 3, 4, 9, 10, 11, 12$},  
\end{eqnarray}  
\begin{eqnarray}  
f_{2}^{(tri, 5)}&=&{e^2\over 4 \pi^2 (m_{H^\pm}^2 - \hat{t})}\left(  
\left[\xi_{3}^2 e_b C_2(m_{H^\pm}^2, 0, \hat{t}, m_{\tilde{t}_1}^2, m_{\tilde{b}_2}^2, m_{\tilde{b}_2}^2)-  
\xi_{3}^2 e_t C_2(m_{H^\pm}^2, 0, \hat{t}, m_{\tilde{t}_1}^2, m_{\tilde{t}_1}^2, m_{\tilde{b}_2}^2) \right]  
\right. \nonumber \\  
&+&\left. \left[\xi_{3}\rightarrow \xi_{2}, m_{\tilde{t}_1}\rightarrow m_{\tilde{t}_2}\right]+  
\left[\xi_{3}\rightarrow \xi_{4}, m_{\tilde{t}_1}\rightarrow m_{\tilde{t}_2}, m_{\tilde{b}_2}\rightarrow m_{\tilde{b}_1}\right]+  
\left[\xi_{3}\rightarrow \xi_{1},  m_{\tilde{b}_2}\rightarrow m_{\tilde{b}_1}\right]  
 \right),  
\end{eqnarray}  
\begin{eqnarray}  
f_{2}^{(tri, 6)}&=&{e^2\over 4 \pi^2 (m_{H^\pm}^2 - \hat{t})}\left(  
\left[ \xi_{3}^2 e_t (C_0(0, m_{H^\pm}^2, \hat{t}, m_{\tilde{t}_1}^2, m_{\tilde{t}_1}^2, m_{\tilde{b}_2}^2) +   
      (C_1+C_2)(m_{H^\pm}^2, 0, \hat{t}, m_{\tilde{b}_2}^2, m_{\tilde{t}_1}^2, m_{\tilde{t}_1}^2 )\right.  
\right. \nonumber \\  
&-&\left.   
\xi_{3}^2 e_b (C_0(0, m_{H^\pm}^2, \hat{t}, m_{\tilde{b}_2}^2, m_{\tilde{b}_2}^2, m_{\tilde{t}_1}^2)+  
       (C_1+C_2)(m_{H^\pm}^2, 0, \hat{t}, m_{\tilde{t}_1}^2, m_{\tilde{b}_2}^2, m_{\tilde{b}_2}^2)\right]  
\nonumber \\  
&+&\left. \left[\xi_{3}\rightarrow \xi_{2}, m_{\tilde{t}_1}\rightarrow m_{\tilde{t}_2}\right]+  
\left[\xi_{3}\rightarrow \xi_{4}, m_{\tilde{t}_1}\rightarrow m_{\tilde{t}_2}, m_{\tilde{b}_2}\rightarrow m_{\tilde{b}_1}\right]+  
\left[\xi_{3}\rightarrow \xi_{1},  m_{\tilde{b}_2}\rightarrow m_{\tilde{b}_1}\right]  
 \right),  
\end{eqnarray}  
\begin{eqnarray}  
f_{2}^{(tri, 8)}=f_{2}^{(tri, 5)}(\hat{u}\leftrightarrow \hat{t}, k_1 \leftrightarrow k_2),  
\end{eqnarray}  
\begin{eqnarray}  
f_{2}^{(tri, 7)}=f_{2}^{(tri, 6)}(\hat{u}\leftrightarrow \hat{t}, k_1 \leftrightarrow k_2),  
\end{eqnarray}  
where $\xi_{j}, (j=1, 2, 3, 4)$ is the coupling constants of the vertex  
$H^+-\tilde{t}-\tilde{b}$, which are given by   
\begin{eqnarray}  
\left( \begin{array}{cc}  
 \xi_{1} & \xi_{3} \\  
        \xi_{4} & \xi_{2}  
 \end{array}\right)=R^{\tilde{t}} {i g \over \sqrt{2} m_w}\left(\begin{array}{cc}  
  -m_w^2 \sin(2 \beta)+m_t^2 cot\beta+m_b^2 tan\beta &  
m_b (\mu+A_b \tan\beta)\\  
m_t (\mu+A_t cot\beta)& m_b m_t (\tan\beta+cot\beta)  
\end{array} \right) (R^{\tilde{b} })^T.  
\end{eqnarray}

$M_{1}^{box}$ is given by  
\begin{eqnarray}  
M_{1}^{box}=\sum_{i=1}^{6} f_{1}^{(box, i)},  
\end{eqnarray}  
where  the form factor $f_1^{(box, i)}$ represents the contributions arising from Feynman   
diagrams with the indices of $i$ in Fig. 4.  
Here   
\begin{eqnarray}  
f_{1}^{(box, 1)}&=&  
-{e^2\over 4 \pi^2}\left( \left[\xi_{4}^2 e_t^2 D_{00}(m_{H^\pm}^2, 0, 0, m_{H^\pm}^2, \hat{t}, \hat{s},   
      m_{\tilde{b}_1}^2, m_{\tilde{t}_2}^2, m_{\tilde{t}_2}^2, m_{\tilde{t}_2}^2) \right.\right. \nonumber \\  
&+&\left.  
D_{00}(m_{H^\pm}^2, 0, 0, m_{H^\pm}^2, \hat{t}, \hat{s},   
 m_{\tilde{t}_2}^2, m_{\tilde{b}_1}^2, m_{\tilde{b}_1}^2, m_{\tilde{b}_1}^2)\right]\nonumber \\  
&+&\left.  
\left[ \xi_{3}\rightarrow \xi_{2}, m_{\tilde{b}_1}\rightarrow m_{\tilde{b}_2} \right]+  
\left[ \xi_{3}\rightarrow \xi_{1}, m_{\tilde{t}_2}\rightarrow m_{\tilde{t}_1} \right]+  
\left[m_{\tilde{t}_2}\rightarrow m_{\tilde{t}_1} ,m_{\tilde{b}_1}\rightarrow m_{\tilde{b}_2} \right]\right),  
\end{eqnarray}  
\begin{eqnarray}  
f_{1}^{(box, 2)}= f_{1}^{(box, 1)}(\hat{t}\leftrightarrow \hat{u}, k_1 \leftrightarrow k_2),  
\end{eqnarray}  
\begin{eqnarray}  
f_{1}^{(box, 3)}&=&  
- { e^2 e_b e_t\over 4 \pi^2}\left(\left[\xi_{3}^2 D_{00}(0, m_{H^\pm}^2, 0, m_{H^\pm}^2, \hat{u}, \hat{t},   
      m_{\tilde{t}_1}^2, m_{\tilde{t}_1}^2, m_{\tilde{b}_2}^2, m_{\tilde{b}_2}^2)\right. \right.\nonumber \\  
&+&\left. D_{00}(0, m_{H^\pm}^2, 0, m_{H^\pm}^2, \hat{u}, \hat{t}, m_{\tilde{b}_2}^2, m_{\tilde{b}_2}^2,   
      m_{\tilde{t}_1}^2, m_{\tilde{t}_1}^2)\right]  
\nonumber \\  
&+&\left.  
\left[ \xi_{3}\rightarrow \xi_{2}, m_{\tilde{t}_1}\rightarrow m_{\tilde{t}_2} \right]+  
\left[ \xi_{3}\rightarrow \xi_{1}, m_{\tilde{b}_2}\rightarrow m_{\tilde{b}_1} \right]+  
\left[m_{\tilde{t}_1}\rightarrow m_{\tilde{t}_2} ,m_{\tilde{b}_2}\rightarrow m_{\tilde{b}_1} \right]\right),  
\end{eqnarray}  
\begin{eqnarray}  
f_{1}^{(box, 4)}&=&\left[ {i \xi_{5} e^2 e_t^2 C_{00}(0, 0, \hat{s}, m_{\tilde{t}_1}^2, m_{\tilde{t}_1}^2, m_{\tilde{t}_1}^2)\over 2 \pi^2}\right]+  
\left[m_{\tilde{t}_1} \rightarrow m_{\tilde{t}_2}, \xi_{5}\rightarrow \xi_{6}\right]\nonumber \\  
&+&  
\left[m_{\tilde{t}_1} \rightarrow m_{\tilde{b}_1}, \xi_{5}\rightarrow \xi_{7}\right]+  
\left[m_{\tilde{t}_1} \rightarrow m_{\tilde{b}_2}, \xi_{5}\rightarrow \xi_{8}\right],  
\end{eqnarray}  
\begin{eqnarray}  
f_{1}^{(box, 5)}&=&{e^2 e_t^2\over 8 \pi^2}\left(\left[\xi_{4}^2 C_0(m_{H^\pm}^2, m_{H^\pm}^2, \hat{s}, m_{\tilde{t}_2}^2, m_{\tilde{b}_1}^2, m_{\tilde{t}_2}^2)  
+  
\xi_{1}^2 C_0(m_{H^\pm}^2, m_{H^\pm}^2, \hat{s}, m_{\tilde{t}_1}^2, m_{\tilde{b}_1}^2, m_{\tilde{t}_1}^2) \right. \right. \nonumber \\  
&+&\left.  
\xi_{2}^2 C_0(m_{H^\pm}^2, m_{H^\pm}^2, \hat{s}, m_{\tilde{t}_2}^2, m_{\tilde{b}_2}^2, m_{\tilde{t}_2}^2)+  
\xi_{3}^2 C_0(m_{H^\pm}^2, m_{H^\pm}^2, \hat{s}, m_{\tilde{t}_1}^2, m_{\tilde{b}_2}^2, m_{\tilde{t}_1}^2)\right]\nonumber \\  
&+&\left.{e_t^2\over 8 \pi^2}\left[m_{\tilde{t}_1}\leftrightarrow m_{\tilde{b}_1}, m_{\tilde{t}_2}\leftrightarrow m_{\tilde{b}_2} \right]  
\right),  
\end{eqnarray}  
\begin{eqnarray}  
f_{1}^{(box, 6)}&=&  
{-i e^2\over 8 \pi^2}\left( \xi_{8} e_b^2 B_0(\hat{s}, m_{\tilde{b}_2}^2, m_{\tilde{b}_2}^2)+  
\xi_{7} e_b^2 B_0(\hat{s}, m_{\tilde{b}_1}^2, m_{\tilde{b}_1}^2)+  
\xi_{6} e_t^2 B_0(\hat{s}, m_{\tilde{t}_2}^2, m_{\tilde{t}_2}^2)\right. \nonumber \\  
&+&\left.  
\xi_{5} e_t^2 B_0(\hat{s}, m_{\tilde{t}_1}^2, m_{\tilde{t}_1}^2)  
\right),  
\end{eqnarray}  
where $\xi_j, (j=5, 6, 7, 8)$ is the coupling constants of the vertexes     
$H^+-H^--\tilde{t}-\tilde{t}$ and $H^+-H^--\tilde{b}-\tilde{b}$, respectively, which are given by  
\begin{eqnarray}  
\left( \begin{array}{cc}  
\xi_{5} & x_1\\  
x_2 & \xi_{6}  
\end{array}\right)=R^{\tilde{t}} i g^2\left(\begin{array}{cc}  
-{\cos (2\beta) (2+sec^2\theta_w)\over 12}-{m_b^2 \tan^2\beta \over  
2 m_w^2}& 0\\  
 0 &{\tan^2\theta_w \cos(2\beta )\over 3}-{m_t^2 cot^2\beta\over 2 m_w^2}  
\end{array} \right) (R^{\tilde{t} } )^T,  
\end{eqnarray}  
\begin{eqnarray}  
\left( \begin{array}{cc}  
\xi_{7} & x_3 \\  
x_4 & \xi_{8}  
\end{array}\right)=R^{\tilde{b}} i g^2\left(\begin{array}{cc}  
 {\cos (2\beta) (4-sec^2\theta_w)\over 12}-{m_t^2 cot^2\beta \over  
2 m_w^2}  
& 0\\  
0 & -{\tan^2\theta_w \cos(2\beta )\over 6}-{m_t^2 \tan^2\beta\over 2 m_w^2}  
 \end{array} \right) (R^{\tilde{b}})^T   
\end{eqnarray}  
with $x_i, (i=1, 2, 3, 4) $ is irrelevant to our calculations.

The form factor $M_{2}^{box}$ is given by  
\begin{eqnarray}  
M_{2}^{box}=\sum_{i=1}^{6} f_{2}^{(box, i)},  
\end{eqnarray}  
where  the form factor $f_2^{(box, i)}$ represents the contributions arising from Feynman   
diagrams with the indices of $i$ in Fig. 4.  
Here   
\begin{eqnarray}  
f_{2}^{(box, 1)}&=&-{e^2\over 4 \pi^2}\left( \left[\xi_{3}^2 (e_t^2+e_b^2) (  
\right.\right. \nonumber \\  
&&(D_0 + 4 D_1 +   
       2 D_2 +   
       2 D_{11}+   
       4 D_{12}+   
       2 D_{13} +   
       D_{22})\nonumber\\  
&&  
(m_{H^\pm}^2, 0, 0, m_{H^\pm}^2, \hat{t}, \hat{s},   
      m_{\tilde{b}_1}^2, m_{\tilde{t}_2}^2, m_{\tilde{t}_2}^2,   
m_{\tilde{t}_2}^2)\nonumber\\  
&+&  
(D_0 + 4 D_1 +   
       2 D_2 +   
       2 D_{11}+   
       4 D_{12}+   
       2 D_{13} +   
       D_{22})\nonumber\\  
&&  
\left.   
 (m_{H^\pm}^2, 0, 0, m_{H^\pm}^2, \hat{t}, \hat{s},   
 m_{\tilde{t}_2}^2, m_{\tilde{b}_1}^2, m_{\tilde{b}_1}^2, m_{\tilde{b}_1}^2)  
 ) \right]\nonumber \\  
&+&  
\left[ \xi_{3}\rightarrow \xi_{2}, m_{\tilde{b}_1}\rightarrow m_{\tilde{b}_2} \right]+  
\left[ \xi_{3}\rightarrow \xi_{1}, m_{\tilde{t}_2}\rightarrow m_{\tilde{t}_1} \right]\nonumber\\  
&+&\left.  
\left[m_{\tilde{t}_2}\rightarrow m_{\tilde{t}_1} ,m_{\tilde{b}_1}\rightarrow m_{\tilde{b}_2} \right]\right),  
\end{eqnarray}  
\begin{eqnarray}  
f_{2}^{(box, 2)}= f_{2}^{(box, 1)}(\hat{t}\leftrightarrow \hat{u},k_1 \leftrightarrow k_2),  
\end{eqnarray}  
\begin{eqnarray}  
f_{2}^{(box, 3)}&=& -{e^2 e_b e_t\over 4 \pi^2}\left( \left[\xi_{3}^2   
(  
(D_2 +D_3 +D_{22} +2D_{23} +D_{33})  
\right.\right. \nonumber \\  
&&(0, m_{H^\pm}^2, 0, m_{H^\pm}^2, \hat{u}, \hat{t},   
        m_{\tilde{t}_1}^2, m_{\tilde{t}_1}^2,   
m_{\tilde{b}_2}^2, m_{\tilde{b}_2}^2)\nonumber\\  
&+&  
(D_2 +D_3 +D_{22} +2D_{23} +D_{33})\nonumber\\  
&&  
\left.   
 (0, m_{H^\pm}^2, 0, m_{H^\pm}^2, \hat{u}, \hat{t}, m_{\tilde{b}_2}^2,   
m_{\tilde{b}_2}^2, m_{\tilde{t}_1}^2, m_{\tilde{t}_1}^2))  
 \right]\nonumber \\  
&+&  
\left[ \xi_{3}\rightarrow \xi_{2}, m_{\tilde{t}_1}\rightarrow m_{\tilde{t}_2} \right]+  
\left[ \xi_{3}\rightarrow \xi_{1}, m_{\tilde{b}_2}\rightarrow m_{\tilde{b}_1} \right]\nonumber\\  
&+&\left.  
\left[m_{\tilde{t}_1}\rightarrow m_{\tilde{t}_2} ,m_{\tilde{b}_2}\rightarrow m_{\tilde{b}_1} \right]\right),  
\end{eqnarray}  
\begin{eqnarray}  
f_{2}^{(box, j)}=0 \mbox{ for $j=4, 5, 6$}.  
\end{eqnarray}  
  


\topmargin 0.5cm

\section*{Figures caption}

\begin{figure} 
\epsfxsize=10 cm
\centerline{\epsffile{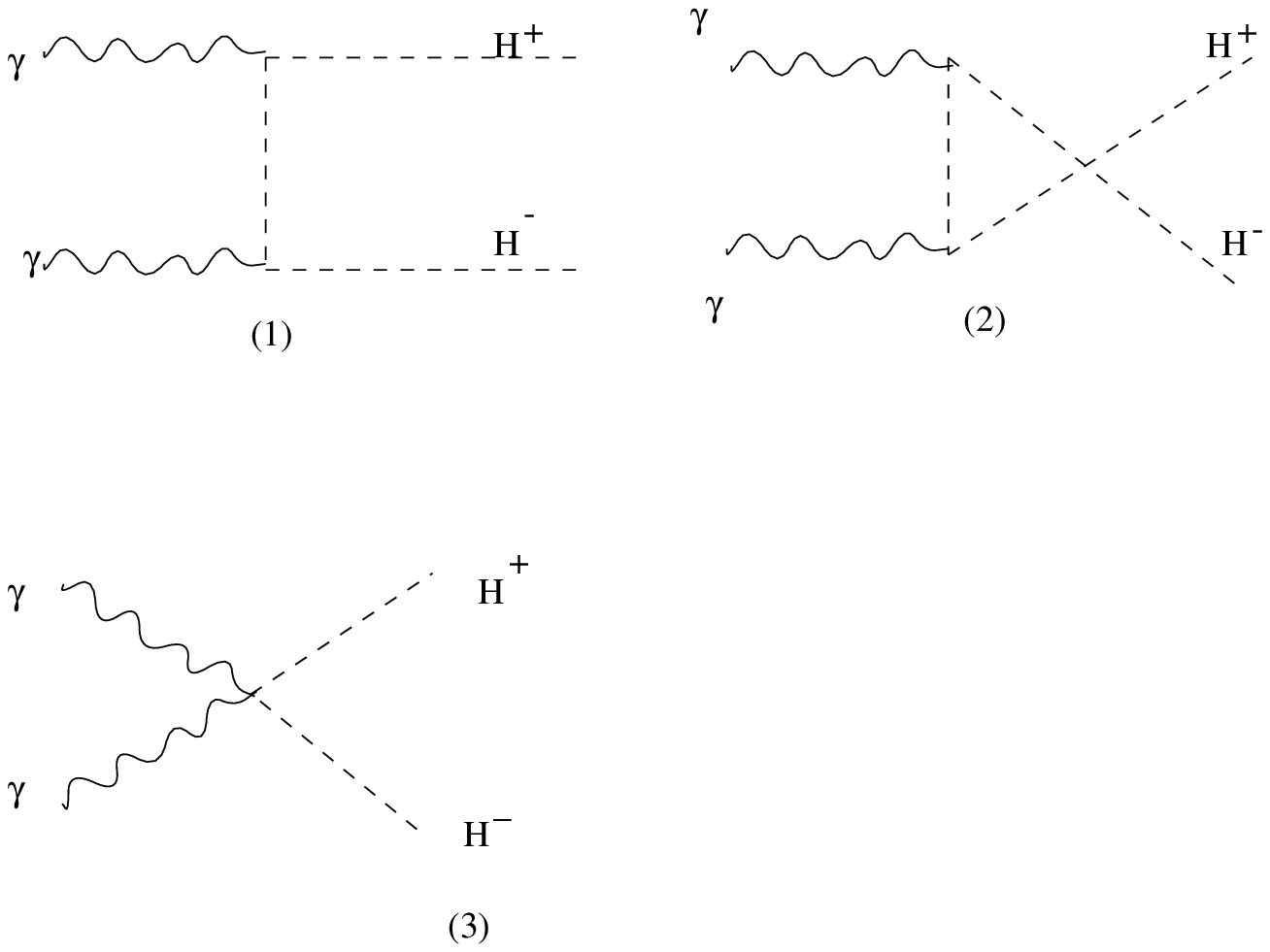}}
\caption[]{ 
} 
\label{fig1}
\end{figure}

 Figure \ref{fig1}: The tree level diagrams for sub-process 
$\gamma\gamma\rightarrow H^+H^-$. In the Feynman diagrams of 
this figure and below, the wavy, 
the dashed and the dotted lines stand for the photon, 
the charged Higgs boson and 
the squark, respectively.

\begin{figure} 
\epsfxsize=10 cm
\centerline{\epsffile{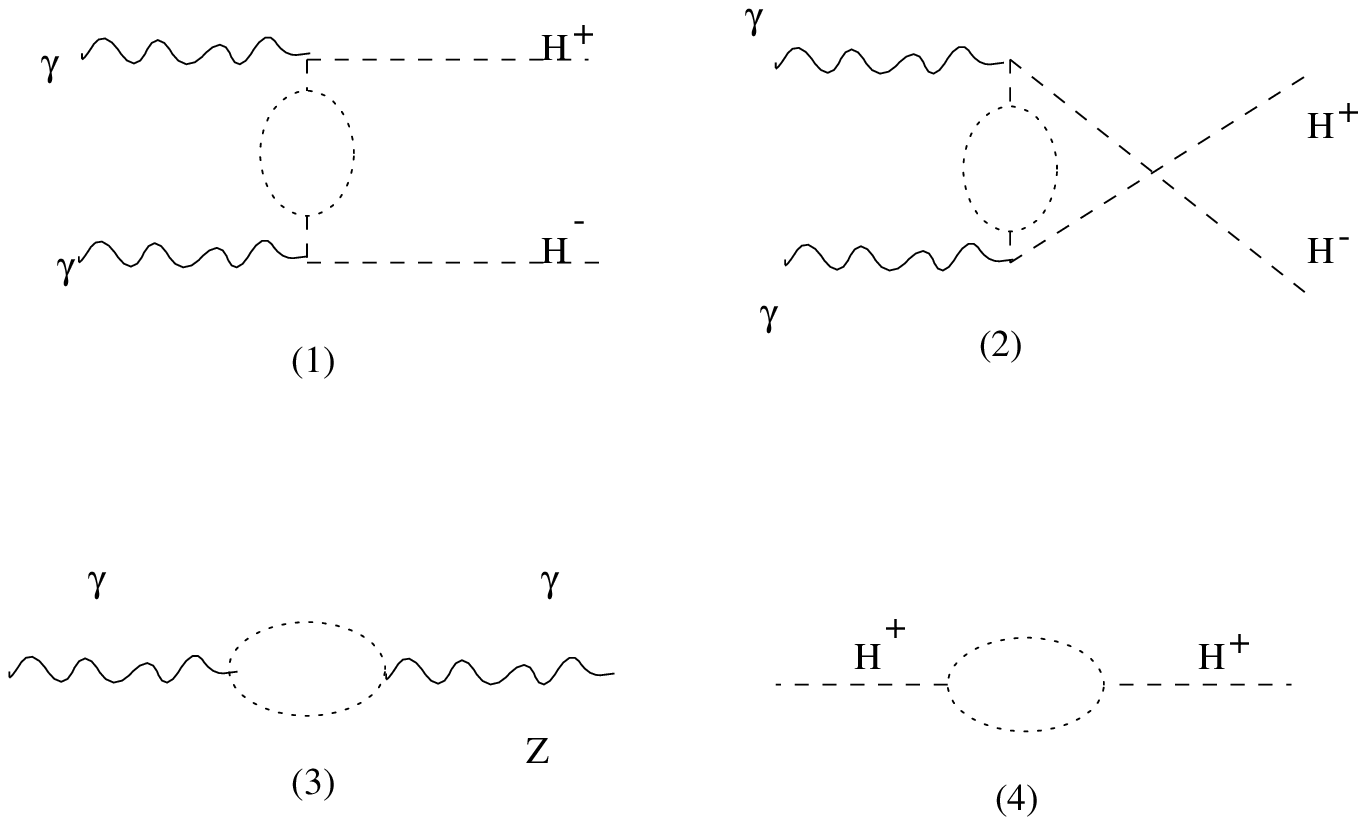}}
\caption[]{ 
} 
\label{fig2}
\end{figure}

 Figure \ref{fig2}: The self-energy correction diagrams of sub-process 
$\gamma\gamma\rightarrow H^+H^-$. 

\begin{figure} 
\epsfxsize=10 cm
\centerline{\epsffile{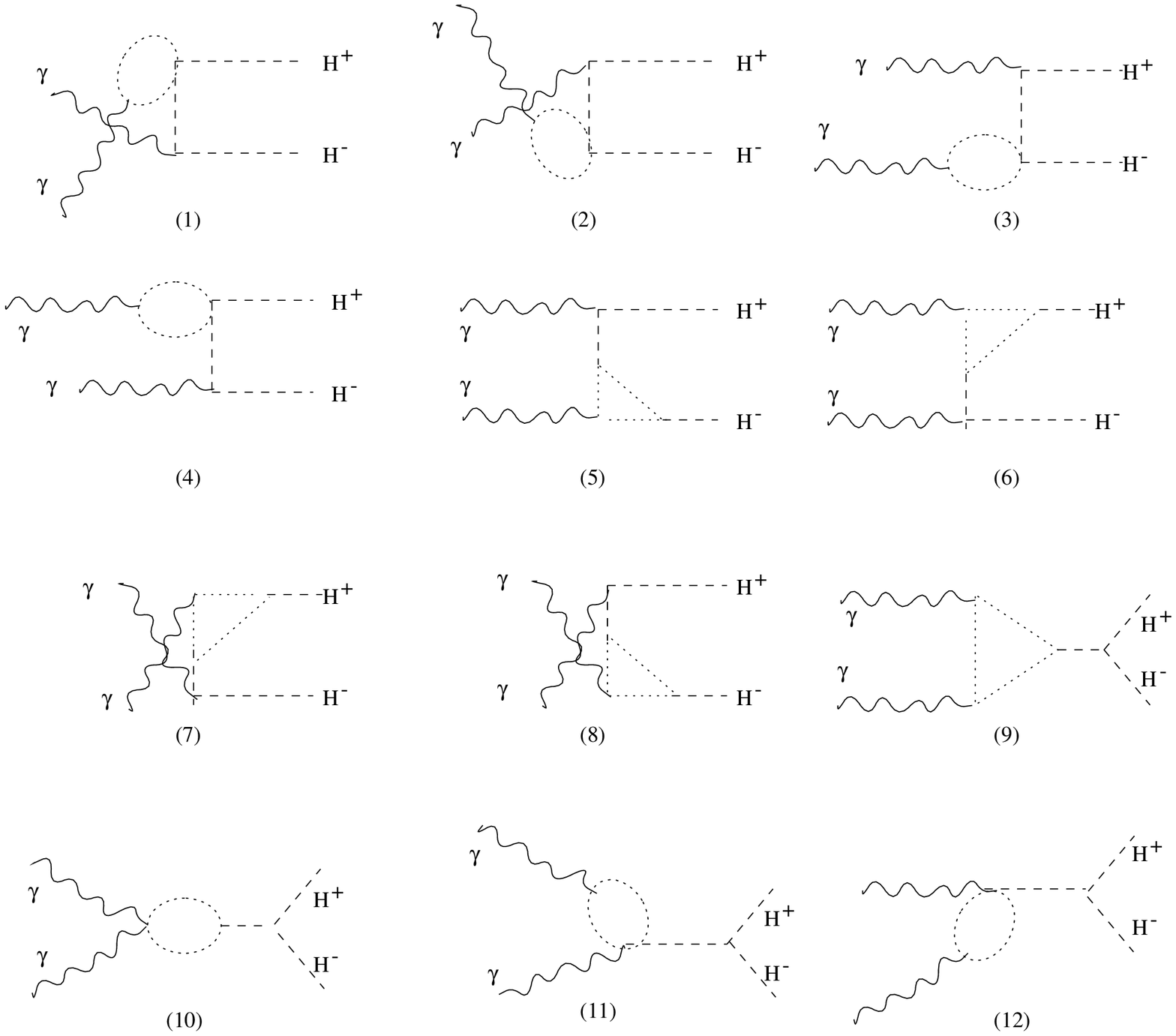}}
\caption[]{
}\label{fig3} 
\end{figure}

Figure \ref{fig3}: The vertex correction diagrams of sub-process 
$\gamma\gamma\rightarrow H^+H^-$.

\begin{figure} 
\epsfxsize=10 cm
\centerline{\epsffile{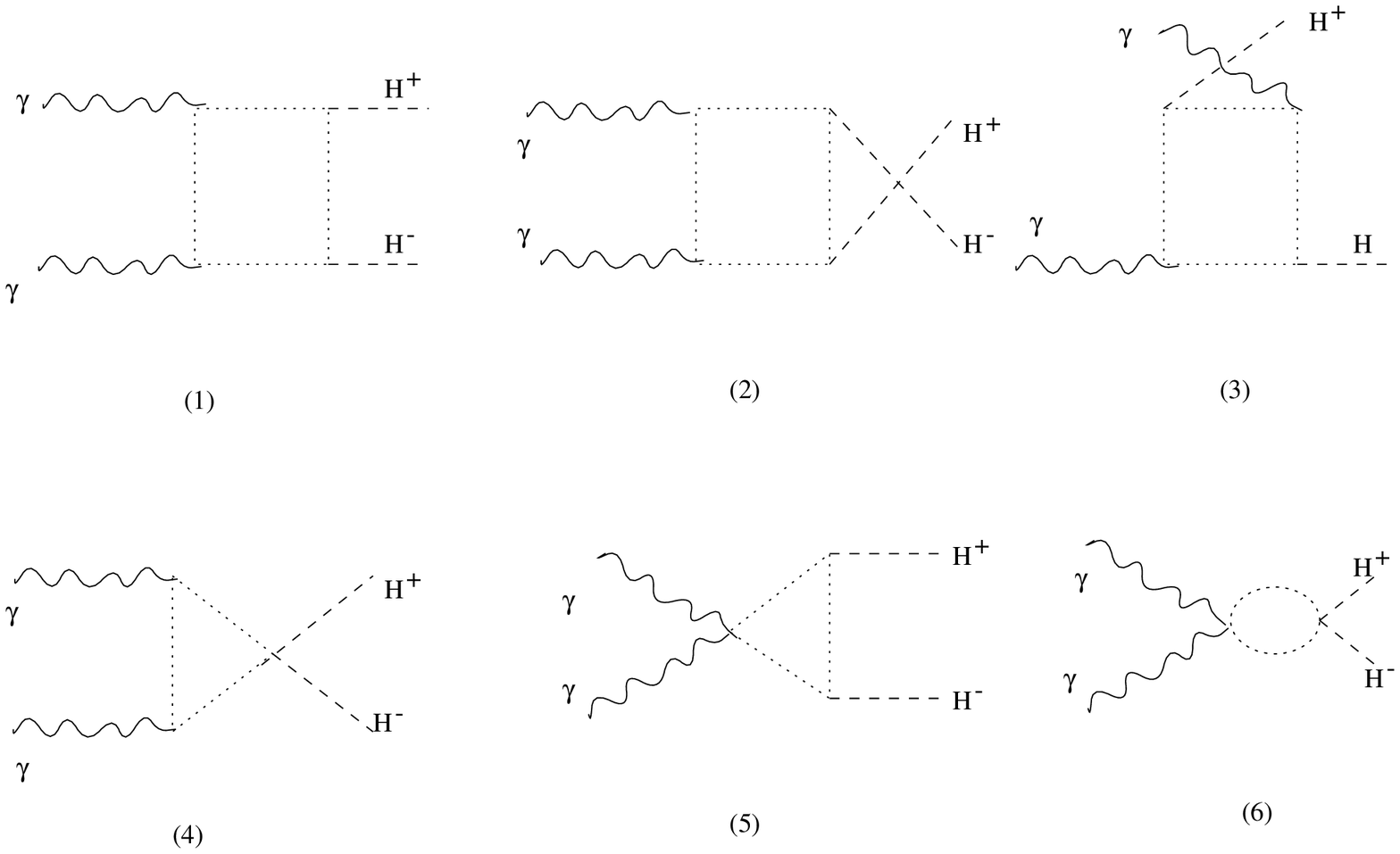}}
\caption[]{
} \label{fig4}
\end{figure}

Figure \ref{fig4}: The box correction diagrams of sub-process
$\gamma\gamma\rightarrow H^+H^-$.

\begin{figure}
\epsfxsize=10 cm
\centerline{\epsffile{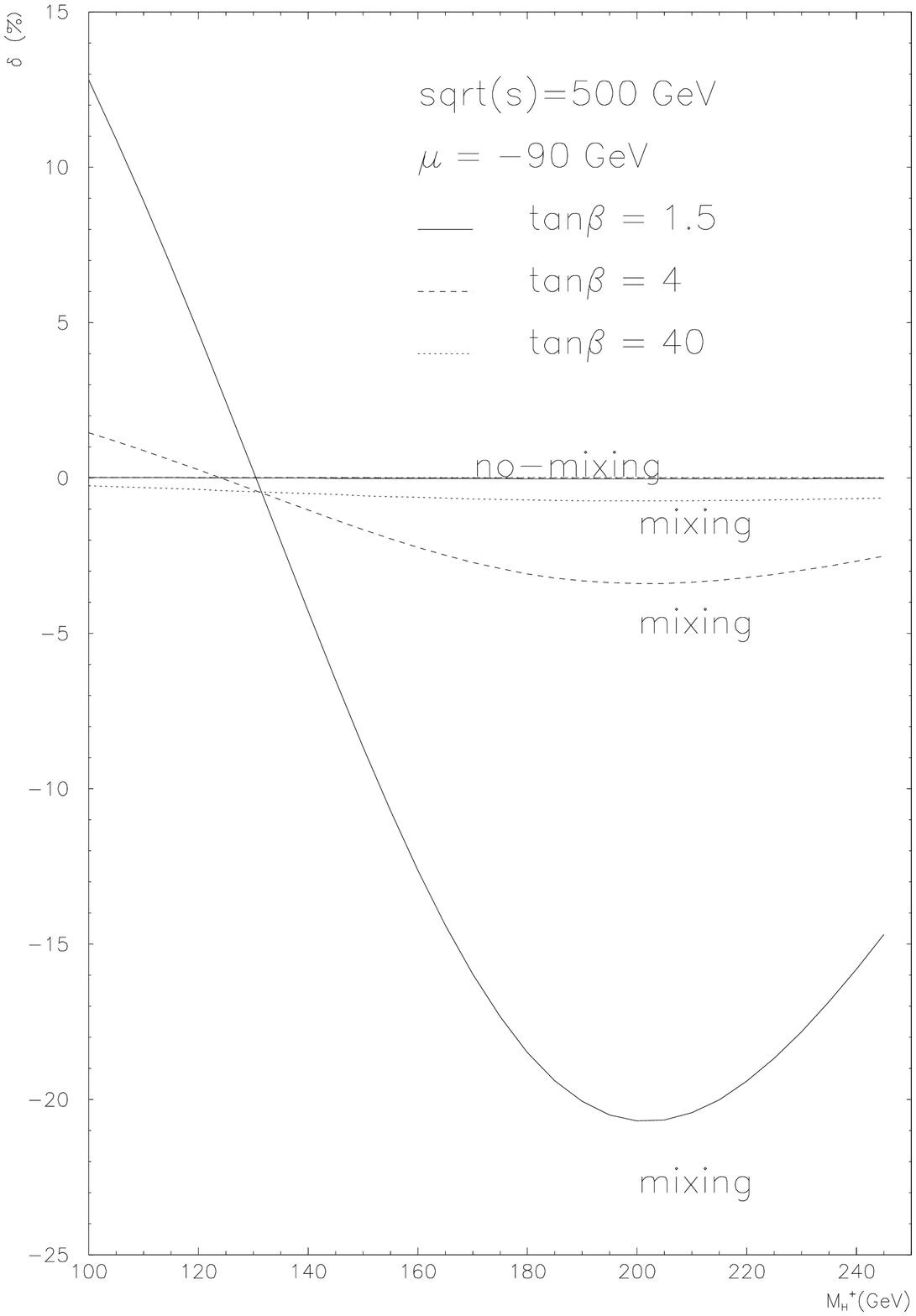}}
\caption[]{
  }
\label{fig5}
\end{figure}

Figure \ref{fig5}: 
Relative corrections as a function of the charged Higgs boson
mass with $\sqrt{\hat{s}}= 500 GeV$, 
where {\em no-mixing } 
represents that there
does not exist mixing between
the left- and right-handed stop, and {\em  mixing } 
represents there is mixing between the stops.
We choose $m_{\tilde{U}}= m_{\tilde{D}}
 =m_{\tilde{Q}}=
 1$ TeV and $\mu = -90$ GeV. In the case of mixing of stops, 
 we choose the value of parameter $A_t$
 which makes the masse of
 the lighter stop equal to $90$ GeV.
 
\begin{figure}
\epsfxsize=10 cm
\centerline{\epsffile{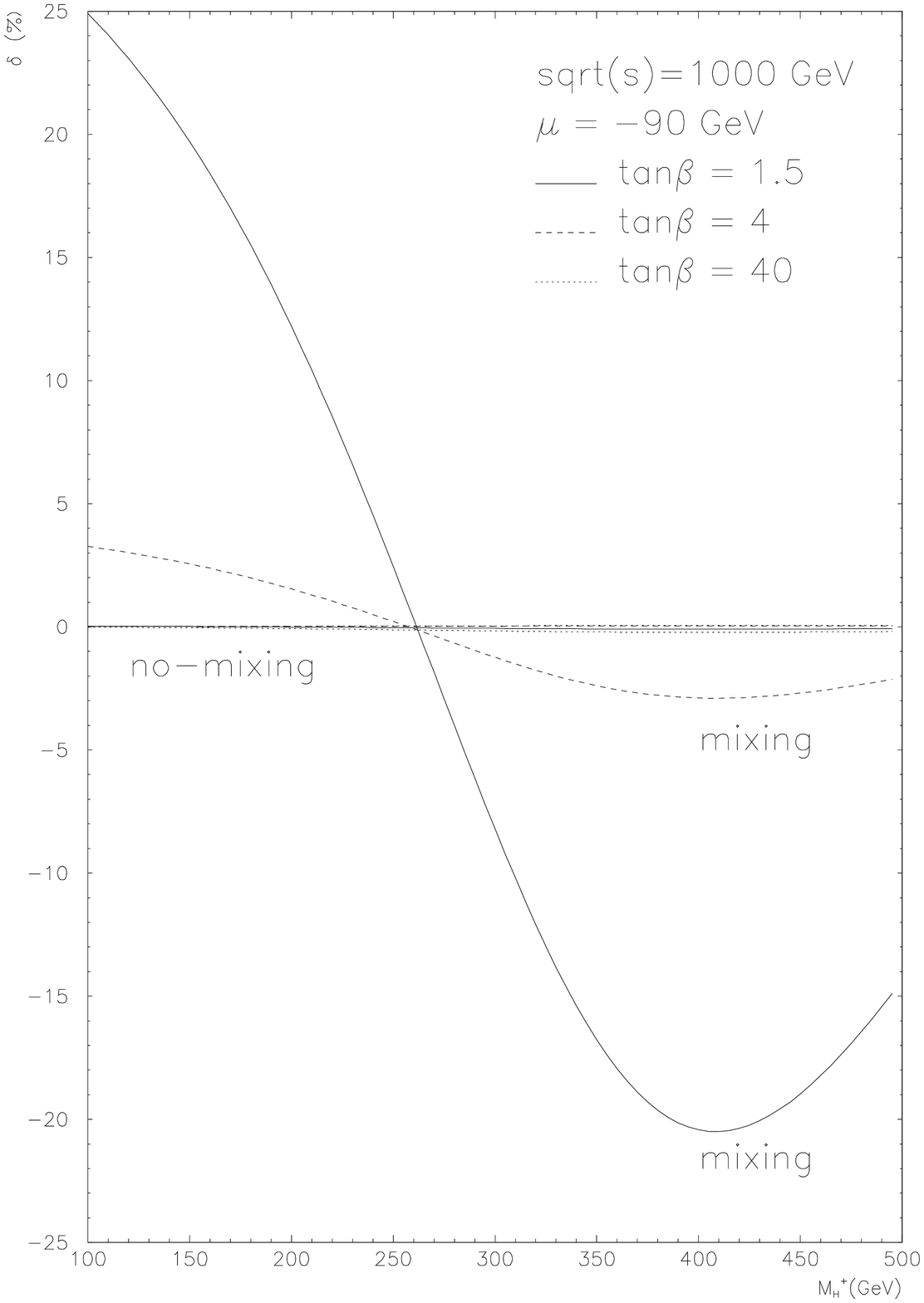}}
\caption[]{
}
\label{fig6}
\end{figure}

 Figure \ref{fig6}: 
 Same with Figure \ref{fig5} but $\sqrt{\hat{s}}= 1000 GeV$.

\begin{figure}
\epsfxsize=10 cm
\centerline{\epsffile{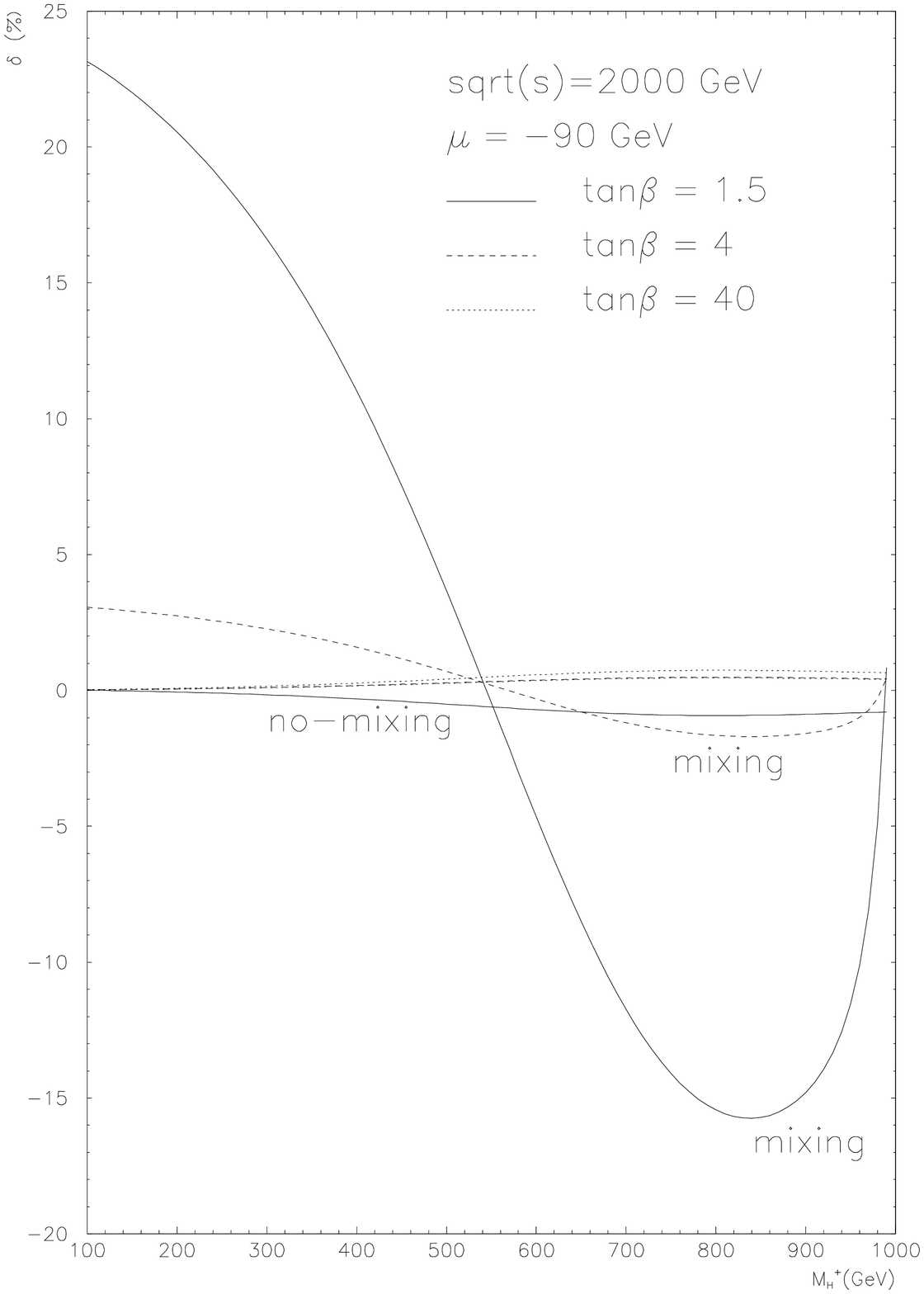}}
\caption[]{ 
}
\label{fig7}
\end{figure}

Figure \ref{fig7}: 
 Same with Figure \ref{fig5} but $\sqrt{\hat{s}}= 2000 GeV$.

\begin{figure}
\epsfxsize=10 cm
\centerline{\epsffile{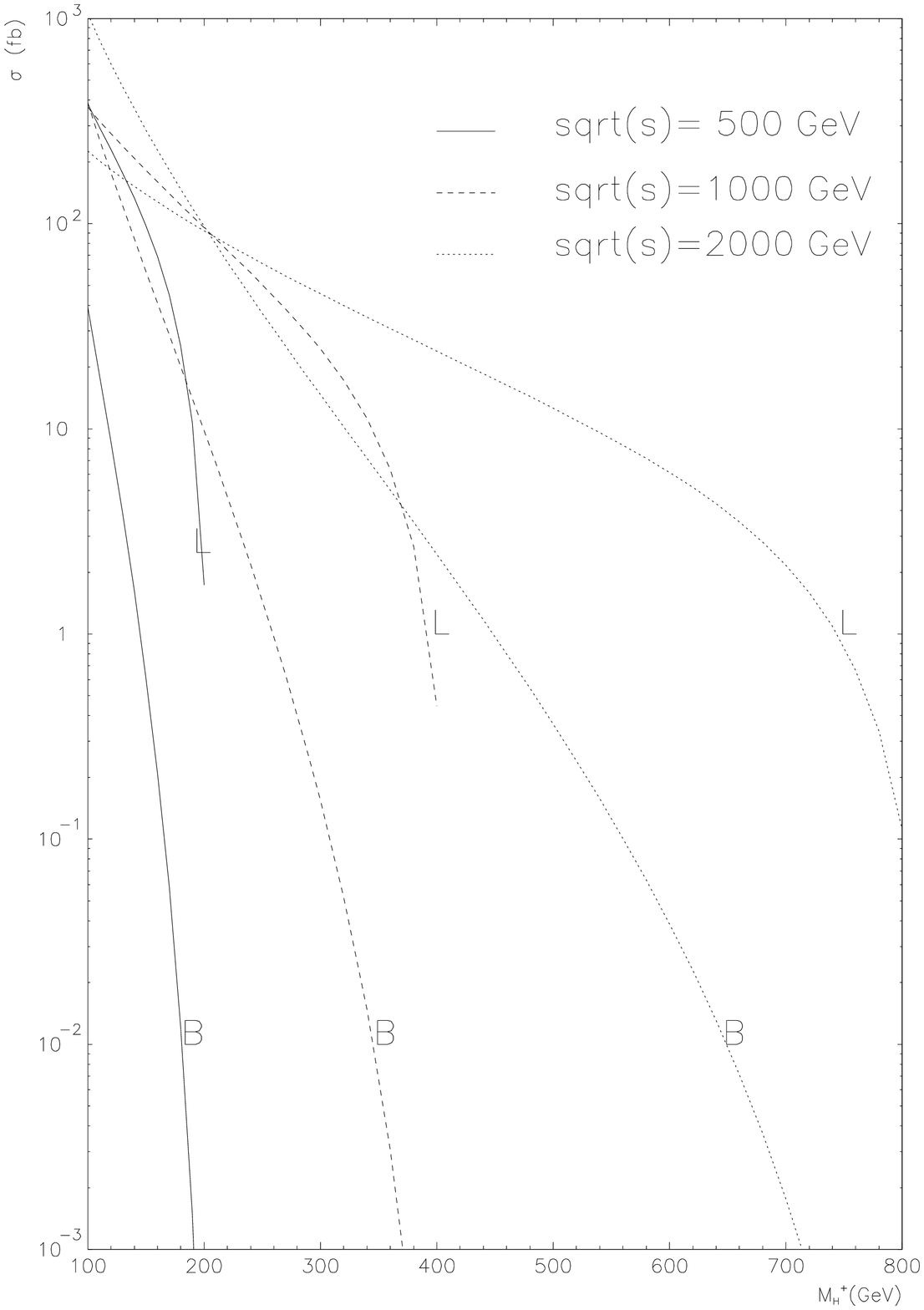}}
\caption[]{ 
 }
\label{fig8}
\end{figure}

Figure \ref{fig8}: 
\  The tree level cross sections for the process $e^+e^- \rightarrow
\gamma\gamma
\rightarrow H^+H^-$ as functions of the charged Higgs boson masses,
where the $e^+e^-$ CMS energies
are $2000$, $1000$ and $500 GeV$, respectively. 
The letters (L) and (B) on the
curves represent the  laser-back scattering
and beamstrahlung photons  modes.

\begin{figure}
\epsfxsize=10 cm
\centerline{\epsffile{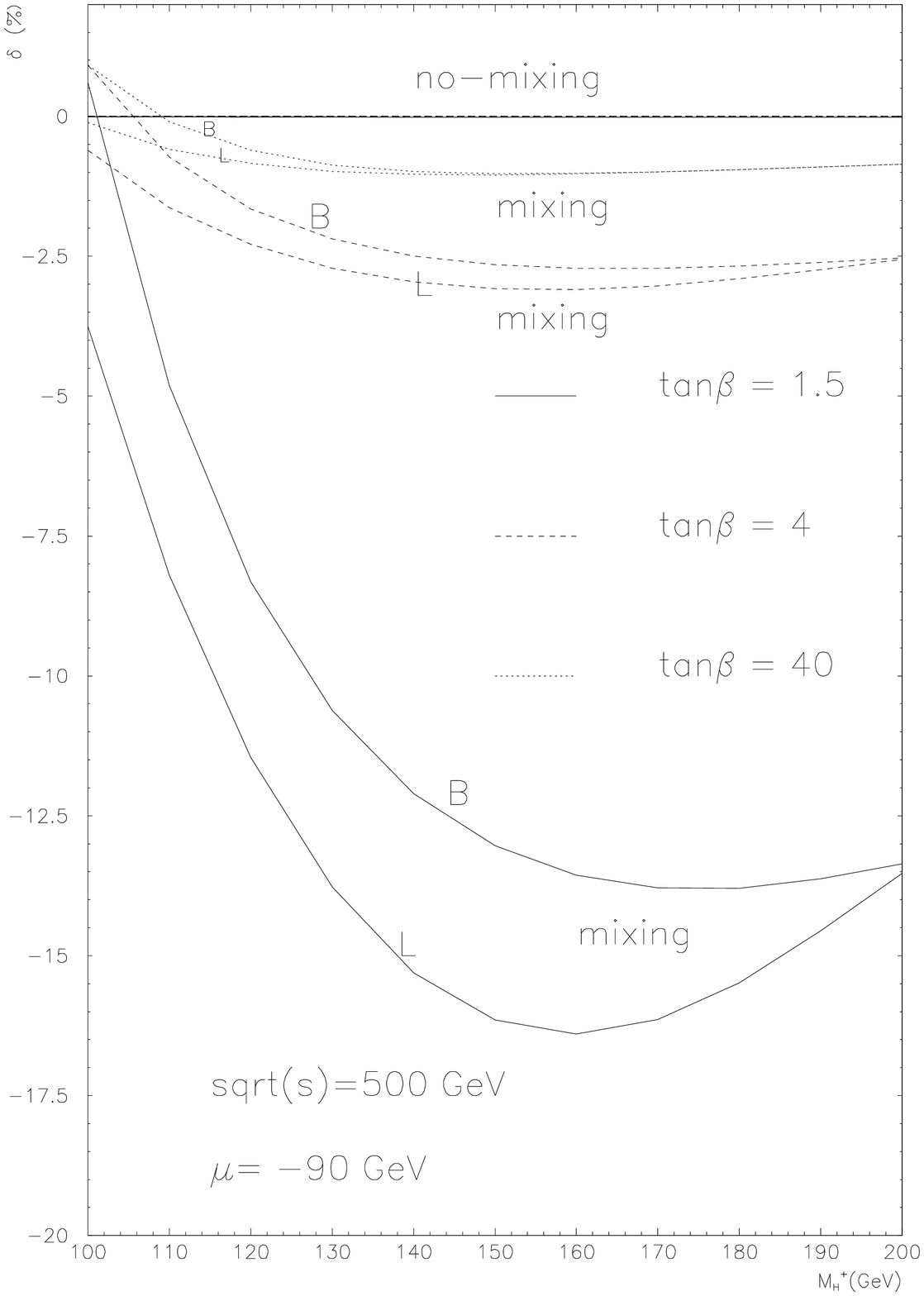}}
\caption[]{
}
\label{fig9}
\end{figure}

Figure \ref{fig9}:  Relative corrections as a function of the charged Higgs boson
mass with $\sqrt{s}= 500 GeV$,
where {\em no-mixing } 
represents that there
does not exist mixing between
the left- and right-handed stop, and {\em  mixing } 
represents there is mixing between the stops.
We choose $m_{\tilde{U}}= m_{\tilde{D}}
 =m_{\tilde{Q}}=
 1$ TeV and $\mu = -90$ GeV. In the case of mixing of stops, 
 we choose the value of parameter $A_t$
 which makes the masse of
 the lighter stop equal to $90 GeV$.
 The letters (L) and (B) on the
curves represent the  laser-back scattering
and beamstrahlung photons  modes.

\begin{figure}
\epsfxsize=10 cm
\centerline{\epsffile{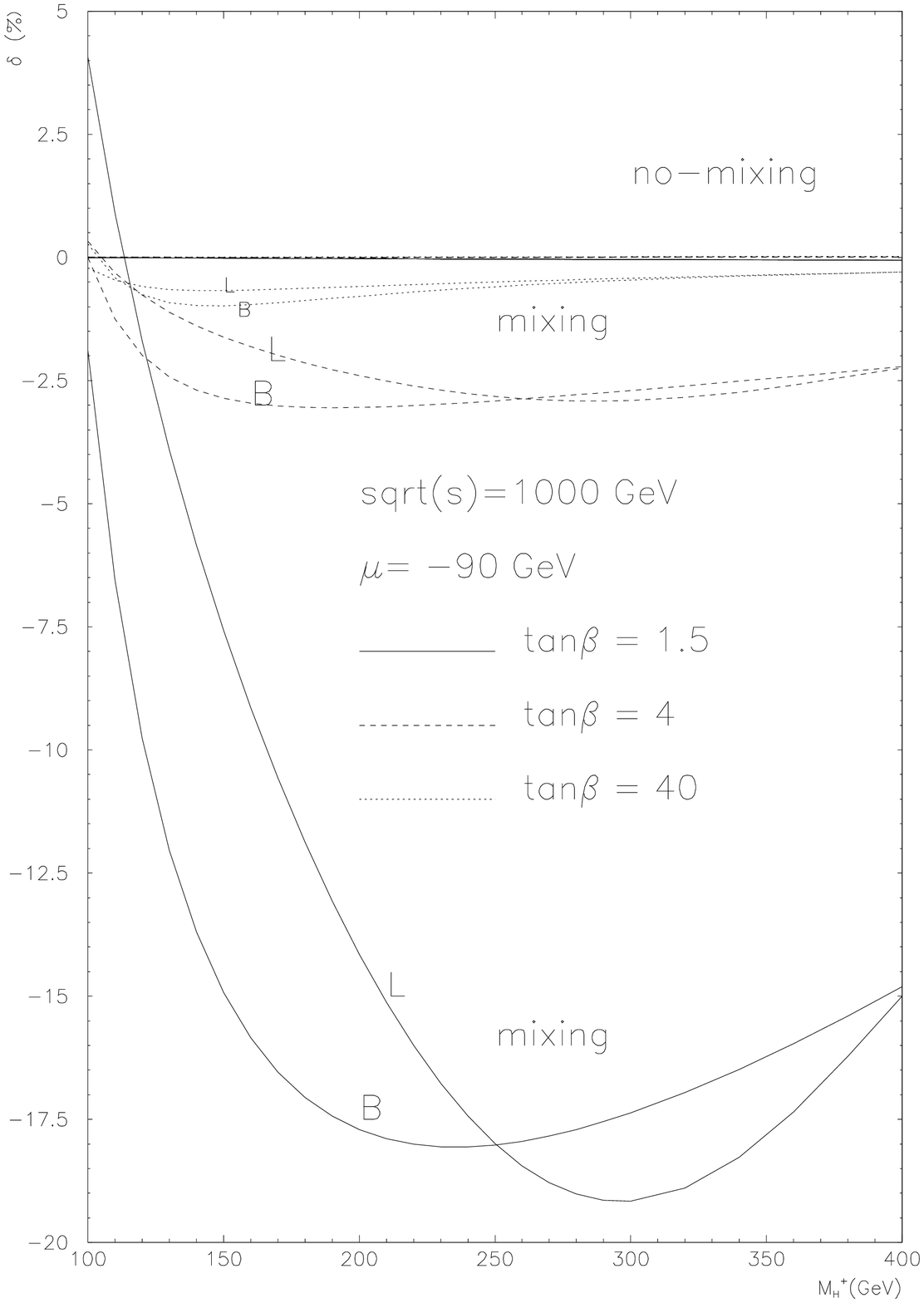}}
\caption[]{
} 
\label{fig10}
\end{figure}

Figure \ref{fig10}:  Same with Figure \ref{fig9} but  the $e^+e^-$ CMS energy
$\sqrt{s}=1000 GeV$.

\begin{figure}
\epsfxsize=10 cm
\centerline{\epsffile{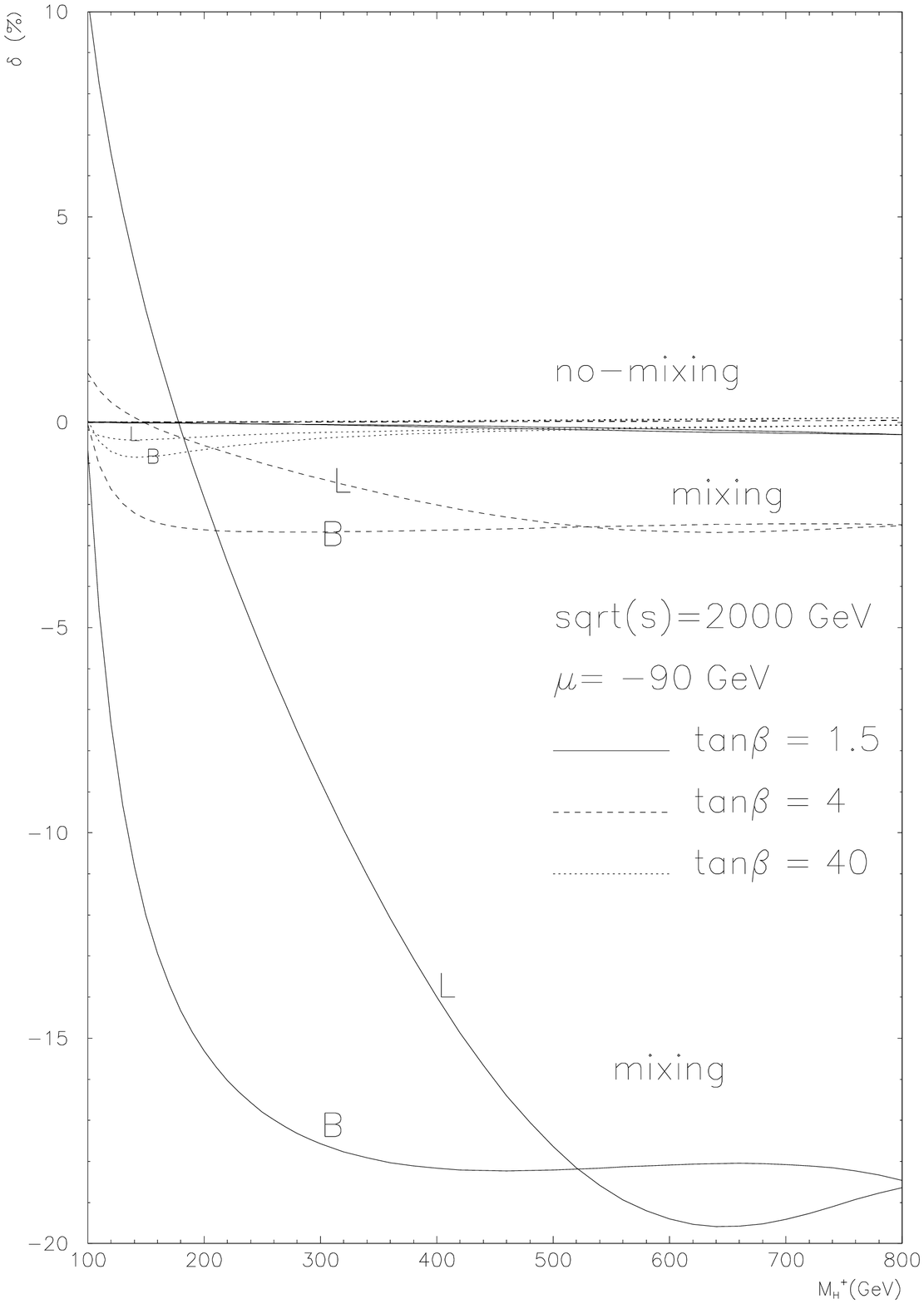}}
\caption[]{
} 
\label{fig11}
\end{figure}

Figure \ref{fig11}:  Same with Figure \ref{fig9} but the $e^+e^-$ CMS energy
$\sqrt{s}=2000 GeV$.

\end{document}